\documentclass[a4paper,11pt]{article}
\usepackage[toc,page]{appendix}
\usepackage{graphicx}
\usepackage{colortbl}
\usepackage{amsmath}
\usepackage{subfigure}
\usepackage{mathtools}
\usepackage[utf8]{inputenc}
\usepackage{float}
\usepackage[normalem]{ulem}
\usepackage{epstopdf}
\usepackage{amsfonts,amsmath,amssymb}
\usepackage{hyperref}

\usepackage{epsfig,multicol,bbm}
\usepackage{color}
\usepackage[dvipsnames]{xcolor}

\definecolor{darkblue}{rgb}{0.0, 0.0, 0.55}
\definecolor{grey}{rgb}{0.57, 0.64, 0.69}
\definecolor{lightbrown}{rgb}{0.71, 0.4, 0.11}

\newcommand{\be}{\begin{equation}}
\newcommand{\ee}{\end{equation}}

\newcommand\fverb{\setbox\pippobox=\hbox\bgroup\verb}
\newcommand\fverbit{\egroup\item[\fbox{\unhbox\pippobox}]}

\newbox\pippobox

\begin{document}
\title{\bf Phase Transition Between Flat Space Cosmology and  Hot Flat Spacetimes in GMMG and EGMG Models}
\author{ M. R. Setare\thanks{Electronic address: rezakord@ipm.ir}\,,\,S. N. Sajadi\thanks{Electronic address: naseh.sajadi@gmail.com}
\\
\small Department of Science, Campus of Bijar, University of Kurdistan, Bijar, Iran\\
}
\maketitle
\begin{abstract}
Flat Space Cosmology (FSC) spacetimes are exact solutions of 3D gravity theories. In this work, we study phase transition between FSC spacetimes and Hot Flat Spacetimes (HFS) in general minimal massive gravity and exotic general massive gravity. We show that similar to topological massive gravity the tunneling occurs between two spacetimes by comparing their free energies. We also obtain the corrections to the Bekenstein-Hawking entropy, and its effect on the phase transition is studied.\\
\end{abstract}

\maketitle
\section{Introduction}
General Relativity (GR) in $(2+1)$-dimensions has features that distinguish it from higher-dimensional theories of
gravity. Due to the absence of local degrees of freedom, GR in three dimensions is a simpler theory. In spite of simplicity, when a negative cosmological constant is included the theory does contain black holes \cite{BTZ}.  Including local degrees of freedom makes the theory closer to its four dimensional counterpart; this can be attained by deforming the theory, giving a mass to the graviton. So, by adding the Einstein-Hilbert action with the parity odd, Lorentz-Chern-Simons term for the Christoffel connection, one can obtain a theory that propagates a massive graviton \cite{TMG1}.
Efforts to figure out, on microscopic terms, the entropy of some of these black holes in
a holographic context cause bulk-boundary clash of these theories, according
to which positivity of the energy of the bulk massive graviton seems to be achieved only at
the cost of introducing a negative central charge within the dual theory. A theory that avoids
this problem is Minimal Massive Gravity (MMG). MMG supplements the field equations of
Topologically Massive Gravity with a symmetric, rank-two tensor containing up to
second derivatives of the metric \cite{MMG}.  A general version of  MMG is the generalized minimal massive gravity (GMMG) theory in which by adding the CS deformation term, and an extra term to pure Einstein gravity with a negative cosmological constant \cite{Setare:2014zea}. This theory is free from negative-energy bulk modes, and also avoids the bulk-boundary unitarity clash. It's been shown that the GMMG model has no ghosts, and this model propagates only two physical modes \cite{Setare:2014zea}. The ADT charges \cite{Setare:2015vea}, horizon fluffs of the theory \cite{Setare:2016qob}, the behaviors and algebras of the symmetries and conserved charges near the horizon of the non-extremal black holes in the context of GMMG has been studied in \cite{Setare:2016jba}. Also,  BMS$_3$ algebra of the flat spacetime solutions of GMMG investigated in \cite{Setare:2017mry}. Following this paper, it is of interest to understand if there exists a  flat space analogue  \cite{Bagchi:2009my} of wellknown Hawking-Page phase transition   between hot AdS and the black hole \cite{Hawking:1982dh} in GMMG and  exotic general massive gravity (EGMG) \cite{Ozkan:2018cxj}, \cite{Afshar:2019npk}. In fact, the main goal of this work is to answer this question.\\
Our paper is organized as follows. In section (\ref{PTGMMG}), we obtain the thermodynamics quantities for the solutions in the framework of GMMG, and by using of first law of thermodynamics obtain the free energies. Then, by comparing the free energy of FSC and HFS, we obtain conditions for phase transition. In subsection (\ref{subloggmmg}), we look at the BMS$_{3}$ central charges of the theory and by using Cardy formula obtain the entropy. Then we obtain the leading order correction to entropy and its effect on the phase transition is studied. We study the phase transition between two space-times and logarithmic correction to entropy in the framework of EGMG from viewpoint of Lagrangian formalism in section (\ref{PTEGMG}) and field equations in appendix (\ref{appa}). We conclude the paper in section (\ref{conclusion}) with a discussion of our results.

\section{Phase Transition in GMMG}\label{PTGMMG}
The Lagrangian of GMMG is obtained by generalization the Lagrangian of generalized massive gravity.
The Lagrangian of GMMG model as \cite{Setare:2014zea}
\begin{equation}\label{eqlag}
L_{GMMG}=-\sigma e.R+\dfrac{\Lambda_{0}}{6}e.e\times e+h.T(\omega)+\dfrac{1}{2\mu}\left(\omega.d\omega+\dfrac{1}{3}\omega.\omega\times \omega \right)-\dfrac{1}{m^2}\left(f.R+\dfrac{1}{2}e.f\times f\right)+\dfrac{\alpha}{2}e.h\times h
\end{equation}
here $m$ is a mass parameter of NMG term \cite{Bergshoeff:2009hq}, $h$ and $f$ are auxiliary one-form fields,
$\Lambda_{0}$ is a cosmological parameter with dimension of mass squared, $\sigma$ sign, $\mu$ is a mass parameter of Lorentz Chern-Simons term, $\alpha$ is a dimensionless parameter, $e$ is a dreibein, $\omega$ is a dualizedspin-connection, $T(\omega)$ and $R(\omega)$ are a Lorentz covariant torsion and a curvature 2-form respectively. The equation for metric is obtained by generalizing field equation of MMG. The field equation is defined as follows \cite{Setare:2014zea}:
\begin{equation}\label{eq1}
\bar{\sigma}G_{\mu \nu}+\Lambda_{0}g_{\mu \nu}+\dfrac{1}{\mu}C_{\mu \nu}+\dfrac{\gamma}{\mu^{2}}J_{\mu \nu}+\dfrac{s}{2m^{2}}K_{\mu \nu}= 0 \, ,
\end{equation}
where
\begin{align}\label{eq2}
&C_{\mu \nu}=\dfrac{1}{2}\epsilon_{\mu}^{\ \alpha \beta}\nabla_{\alpha}(R_{\beta \nu}-\dfrac{1}{4}g_{\nu \beta}R),\hspace{0.1cm}
J_{\mu \nu}=R_{\mu \alpha}R^{\alpha}_{\nu}-\dfrac{3}{4}R R_{\mu \nu}-\dfrac{1}{2}g_{\mu \nu}\left(R^{\alpha \beta}R_{\alpha \beta}-\dfrac{5}{8}R^{2}\right),\nonumber\\
K_{\mu \nu}=&-\dfrac{1}{2}\nabla^{2}R g_{\mu \nu}-\dfrac{1}{2}\nabla_{\mu}\nabla_{\nu} R+2\nabla^{2}R_{\mu \nu}+4 R_{\mu a \nu b}R^{a b}-
\dfrac{3}{2}R R_{\mu \nu}-R_{\alpha \beta}R^{\alpha \beta}g_{\mu \nu}+\dfrac{3}{8}R^{2}g_{\mu \nu}
\end{align}
 and $s$ is sign, $\gamma$, $\bar{\sigma}$ and $\Lambda_{0}$  are the parameters which defined in terms of cosmological constant $\Lambda$, $m$, $\mu$, and the sign of Einstein-Hilbert term. Symmetric tensors $J_{\mu \nu}$ and $K_{\mu \nu}$ are coming from MMG and NMG parts respectively.\\
 We consider following metric \cite{Setare:2017mry}, \cite{Detournay:2016sfv}
\begin{align}\label{cosmic}
ds^2=&\mathcal{M}(\phi)du^2+2\mathcal{N}(u,\phi)du d\phi -2e^{\mathcal{A}(\phi)}dr du-2\beta(u,\phi) e^{\mathcal{A}(\phi)} dr d\phi+\nonumber\\
&\left(e^{2\mathcal{A}(\phi)}r^2+\beta(u,\phi)(2\mathcal{N}(u,\phi)-\beta(u,\phi) \mathcal{M}(\phi))\right)d\phi^2
\end{align}
where $\mathcal{N}=\mathcal{L}+u/2\partial_{\phi}\mathcal{M}$ and $\beta=\mathcal{B}+u\partial_{\phi}\mathcal{A}$, and $\mathcal{M}(\phi)$,
$\mathcal{A}(\phi)$, $\mathcal{B}(\phi)$, $\mathcal{L}(\phi)$ are arbitrary functions of $\phi$. From now on, we consider the case $\mathcal{N}=\mathcal{L}=const$, $\beta=\mathcal{B}=const$, $\mathcal{A}=const$ and $\mathcal{M}=const$. So, the metric (\ref{cosmic}) becomes
\begin{align}\label{cosmic1}
ds^2=\mathcal{M}du^2+2\mathcal{L}du d\phi -2e^{\mathcal{A}}dr du-2\mathcal{B} e^{\mathcal{A}} dr d\phi+
\left(e^{2\mathcal{A}}r^2+\mathcal{B}(2\mathcal{L}-\mathcal{B}\mathcal{M})\right)d\phi^2.
\end{align}
This metric is a flat space limit of BTZ black hole by scaling $l\rightarrow \infty$. In the limit, the outer horizon $r_+$ is pushed to infinity, i.e, $r_+=-l\sqrt{\mathcal{M}} $ and the inner horizon becomes a cosmological horizon , i.e, $r_-=\mathcal{L}/\sqrt{\mathcal{M}}$ \cite{Barnich:2012xq}.
 For $\mathcal{M}=-1, \mathcal{L}=\mathcal{A}=\mathcal{\beta}=0$, one can obtain a three dimensional flat spacetime in retarded Bondi coordinates as follows
\begin{equation}\label{Bmetric}
ds^{2}=-du^2-2du dr+r^2 d\phi^2.
\end{equation}
where $du=dt-dr$ and $\phi\sim \phi+2\pi$ and is also a solution for the field equations (\ref{eq1}) in $\Lambda_{0}=0$.
To study the phase transition, we continue by comparing the free energy of the spacetimes. Spacetimes with smaller free energy is preferred thermodynamically.
The Euclidean signature version of flat space allows us to
introduce a finite temperature by periodically identifying
the Euclidean time. To go to the Euclidean signature, we need $u=-iu_{E}$, $r=-ir_{E}$ and $\phi=-i\phi_{E}$ in the metrics (\ref{Bmetric}) and (\ref{cosmic}) as follows \cite{Bagchi:2013lma}
\begin{equation}
ds_{E}^2=du_{E}^2+2du_{E}dr_{E}+r_{E}^2d\phi_{E}^2.
\end{equation}
and
\begin{align}
ds_{E}^2=&-\mathcal{M}du_{E}^2-2\mathcal{L}du_{E} d\phi_{E} +2e^{\mathcal{A}}dr_{E} du_{E}+2\beta e^{\mathcal{A}} dr_{E} d\phi_{E}+\left(e^{2\mathcal{A}}r_{E}^2+\mathcal{B}(2\mathcal{L}-\mathcal{B} \mathcal{M})\right)d\phi_{E}^2
\end{align}
where $u_{E} \sim u_{E}+\beta$, $\phi_E \sim \phi_{E}+\beta \Omega$ and $\beta=T^{-1}$.
The temperature by using of surface gravity on cosmological horizon is \cite{Setare:2017mry}
\begin{equation}\label{eqtemp}
T=\dfrac{\kappa}{2\pi}=\left.\dfrac{1}{2\pi}\sqrt{-\dfrac{1}{2}\nabla^{\mu}\xi^{\nu}\nabla_{\mu}\xi_{\nu}}\right\vert_{r_{H}}=\dfrac{\mathcal{M}^{\frac{3}{2}}}{2\pi \mathcal{L}},
\end{equation}
here $\xi^{\mu}=\delta^{\mu}_{u}+\Omega\delta^{\mu}_{\phi}$ and the angular velocity $\Omega$ of the horizon is given by \cite{Setare:2017mry}
\begin{equation}\label{eqomega}
\vert\Omega\vert=\left.-\dfrac{g_{u\phi}}{g_{\phi \phi}}\right\vert_{r_{H}}=\dfrac{\mathcal{M}}{\mathcal{L}},
\end{equation}
where cosmological horizon is
\begin{equation}\label{eqevr}
r_{H}=\dfrac{e^{-\mathcal{A}}}{\sqrt{\mathcal{M}}}\vert\mathcal{L}-\mathcal{M}\mathcal{B}\vert.
\end{equation}
 As can be seen from  (\ref{eqevr}) in the case of $\phi$ dependent metric coeffecent, the location of the horizon radius in an angle dependent way changes and thus describe non spherically symmetric cosmological solutions.\\
The energy, obtained as the conserved charge associated with future-directed time translations generated
by the Killing vector $ \bar{\xi}^{\mu}=\delta^{\mu}_{u} $ using quasi-local conserved charges, reads \cite{Setare:2017mry},\cite{Setare:2017wuj}
 \begin{equation}\label{eqmasssta}
 E={Q}^{\mu}(\bar{\xi})=-\dfrac{1}{8G}\left(\bar{\sigma}+\dfrac{\alpha H}{\mu}+\dfrac{F}{m^2}\right)\mathcal{M},
 \end{equation}
 where $F$ and $H$ are some constants. By considering $\bar{\xi}^{\mu}=-\delta^{\mu}_{\phi} $, angular momentum likewise obtained as follows \cite{Setare:2017mry}
 \begin{equation}\label{eqJ11}
 J=\dfrac{1}{4G}\left[-\left(\bar{\sigma}+\dfrac{\alpha H}{\mu}+\dfrac{F}{m^2}\right)\mathcal{L}-\dfrac{\mathcal{M}}{2\mu}\right],
 \end{equation}
while entropy for theories with a gravitational Chern-Simons term is \cite{Setare:2017mry}
\begin{equation}\label{eqs1}
S=\dfrac{1}{4G}\int \dfrac{d\phi}{\sqrt{g_{\phi \phi}}}\left[-\left(\bar{\sigma}+\dfrac{\alpha H}{\mu}+\dfrac{F}{m^2}\right)g_{\phi \phi}+\dfrac{\Omega_{\phi \phi}}{2\mu}\right],
\end{equation}
where we have
\begin{equation}
\left. g_{\phi \phi}\right\vert_{r_{H}}=\dfrac{\mathcal{L}^2}{\mathcal{M}},\;\;\;\left. \Omega_{\phi \phi}\right\vert_{r_{H}}=\mathcal{L}.
\end{equation}
Then the entropy of FSC becomes
\begin{equation}\label{eqs2}
S=-\dfrac{\pi}{2G}\left[\left(\bar{\sigma}+\dfrac{\alpha H}{\mu}+\dfrac{F}{m^2}\right)\dfrac{\mathcal{L}}{\sqrt{\mathcal{M}}}+\dfrac{\sqrt{\mathcal{M}}}{\mu}\right]=-\dfrac{\pi^2 T}{G\Omega^2}\left[\bar{\sigma}+\dfrac{\alpha H}{\mu}+\dfrac{F}{m^2}+\dfrac{\Omega}{\mu}\right],
\end{equation}
 which agree with the conserved charges of TMG for the solution in the limit $m^2\rightarrow \infty, \bar{\sigma}=-1,\alpha=0$ \cite{Detournay:2016sfv}. Also, all thermodynamic variables of the cosmological solutions are precisely the limit $l\rightarrow \infty $ of the variables of the inner horizon of the BTZ black hole \cite{Barnich:2012xq}.\\
By integration the first law of thermodynamics
\begin{equation}\label{eqflaw}
dF=-S dT+J d\Omega,
\end{equation}
and using equations (\ref{eqtemp})-(\ref{eqJ11}) and (\ref{eqs2}), one can obtain the free energy as follows
\begin{equation}\label{eqfec}
F_{FSC}=-\dfrac{\mathcal{M}}{8G}\left[\bar{\sigma}+\dfrac{\alpha H}{\mu}+\dfrac{F}{m^2}+\dfrac{\mathcal{M}}{\mu \mathcal{L}}\right]=-\dfrac{\pi^2 T^2}{2G \Omega^2}\left[\bar{\sigma}+\dfrac{\alpha H}{\mu}+\dfrac{F}{m^2}+\dfrac{\Omega}{\mu}\right].
\end{equation}
On the other hand one can obtain the free energy of hot flat space by puting $\mathcal{M}=1,\Omega=0$ as follows
\begin{equation}\label{eqff}
F_{HFS}=-\dfrac{1}{8G}\left(\bar{\sigma}+\dfrac{\alpha H}{\mu}+\dfrac{F}{m^2}\right).
\end{equation}
Comparing the free energies (\ref{eqfec}) and (\ref{eqff}) give us that the phase transition between two spacetimes accurs and therefore, the conditions for this phase transition are
\begin{align}\label{eqphasec}
F_{FSC}<F_{HFS}\;\;\rightarrow\;\;&\mathcal{M}\left(1+\dfrac{\Omega}{\mu\left(\bar{\sigma}+\dfrac{\alpha H}{\mu}+\dfrac{F}{m^2}\right)}\right)<1,\;\;\;\;\;\textit{FSC is preferred},\nonumber\\
F_{FSC}=F_{HFS}\;\;\rightarrow\;\;&\mathcal{M}\left(1+\dfrac{\Omega}{\mu\left(\bar{\sigma}+\dfrac{\alpha H}{\mu}+\dfrac{F}{m^2}\right)}\right)=1\;\;\;\;\;\textit{FSC and HFS coexist},\nonumber\\
F_{FSC}>F_{HFS}\;\;\rightarrow\;\;&\mathcal{M}\left(1+\dfrac{\Omega}{\mu\left(\bar{\sigma}+\dfrac{\alpha H}{\mu}+\dfrac{F}{m^2}\right)}\right)>1\;\;\;\;\;\textit{HFS is preferred}.
\end{align}
As can be seen from above, by assuming $\mu>0, \bar{\sigma}+\alpha H/\mu+F/m^2\neq 0$ at sufficiently large $\mathcal{M}$, $T$ and small $\Omega$ the preferred spacetime is FSC. In the case of $\bar{\sigma}+\alpha H/\mu+F/m^2= 0$, the phase transition does not occur.  By using equation (\ref{eqtemp}) and (\ref{eqphasec}), the critical temprature where the phase transition occurs is
\begin{equation}
T_{c}=\dfrac{\Omega}{2\pi}\dfrac{1}{\sqrt{1+\dfrac{\Omega}{\mu \left(\bar{\sigma}+\dfrac{\alpha H}{\mu}+\dfrac{F}{m^2}\right)}}}.
\end{equation}
Thus, below $T_c$ HFS is the dominant spacetime and at critical temperature $T_c$, HFS tunnels into FSC. Conversely, at sufficiently high angular velocity, HFS is more stable and FSC falls apart and tunnels to HFS.\\
Also, by using of heat capacity one can look at the stability of FSC.
From (\ref{eqs2}) with fixed $J$ heat capacity becomes
\begin{equation}\label{eqc21}
C=\left. T\dfrac{\partial S}{\partial T}\right \vert_J=S,
\end{equation}
therefore, from positivity of entropy ($\bar{\sigma}+\alpha H/\mu+F/m^2+\Omega/\mu <0$) one can find the system is localy stable.\\
If we consider the conserved charges of \cite{Setare:2015vea}, which has been obtained by linearizing field equations (\ref{eq1}) and using ADT method, the results for phase transition are the same as TMG \cite{Bagchi:2013lma}.

\subsection{Logarithmic Correction to Entropy}\label{subloggmmg}

In the following we obtain the entropy from viewpoint of CFT.
The  central charges of the CFT dual theory are \cite{Setare:2014zea}
\begin{equation}
c_{\pm}=-\dfrac{3l}{2G}\left(\bar{\sigma}+\dfrac{\alpha H}{\mu}+\dfrac{F}{m^2}\pm \dfrac{1}{\mu l}\right),
\end{equation}
then corresponding BMS$_3$ central charges become
\begin{equation}\label{eqcentral}
c_{L}=\lim_{l\rightarrow \infty} (c_{+}-c_{-})=-\dfrac{3}{\mu G},\;\;\;\;\; c_{M}=\lim_{l\rightarrow \infty}\dfrac{1}{l}(c_{+}+c_{-})=-\dfrac{3}{G}\left(\bar{\sigma}+\dfrac{\alpha H}{\mu}+\dfrac{F}{m^2}\right).
\end{equation}

In the AdS$_3$/CFT$_2$ correspondence, the mass $M$ and angular momentum $J$ of the BTZ black holes are mapped to the conformal weights ($h,\bar{h}$) of the dual CFT by the relations \cite{Bagchi:2013qva}
\begin{equation}
h=\dfrac{lM+J}{2}+\dfrac{c}{24},\;\;\;\;\bar{h}=\dfrac{lM-J}{2}+\dfrac{\bar{c}}{24}
\end{equation}
where $c=\bar{c}=3l/2G$. This implies that for the FSC by using (\ref{eqcentral}), the mass and angular momentum are mapped to \cite{Bagchi:2013qva}
\begin{align}\label{eqcw}
&h_{M}=\lim_{l\rightarrow \infty}\dfrac{h+\bar{h}}{l}=M+\dfrac{c_{M}}{24}=-\dfrac{1}{8G}\left(\bar{\sigma}+\dfrac{\alpha H}{\mu}+\dfrac{F}{m^2}\right)\left(\mathcal{M}+1\right),\nonumber\\
&h_{L}=\lim_{l\rightarrow \infty}(h-\bar{h})=J+\dfrac{c_{L}}{24}=-\dfrac{1}{4G}\left[\left(\bar{\sigma}+\dfrac{\alpha H}{\mu}+\dfrac{F}{m^2}\right)\mathcal{L}+\dfrac{\mathcal{M}}{2\mu}+\dfrac{1}{\mu}\right].
\end{align}
The logarithmic correction to the entropy  \cite{Carlip:2000nv}, \cite{Bagchi:2013qva}
\begin{equation}\label{eqents}
S=2\pi\left(c_{L}\sqrt{\dfrac{h_{M}}{6c_{M}}}+h_{L}\sqrt{\dfrac{c_{M}}{6h_{M}}}\right)-\dfrac{3}{2}\log\left(\dfrac{2h_{M}}{c_{M}}\right)+...=S^{(0)}+S^{(1)}+...
\end{equation}
for the case of GMMG by using of (\ref{eqcentral}) and (\ref{eqcw}) we get
\begin{equation}\label{eqss}
S^{(0)}=-\dfrac{\pi}{2G\mu}\dfrac{1}{\sqrt{\mathcal{M}+1}}\left[\mathcal{M}+1+\left(\bar{\sigma}+\dfrac{\alpha H}{\mu}+\dfrac{F}{m^2}\right)\mathcal{L}\mu\right],
\end{equation}
in the case of $\mathcal{M}, \mathcal{L}\gg 1$, the results (\ref{eqss}) and (\ref{eqs2}) are the same.
The correction term to entropy becomes
\begin{equation}
S^{(1)}=-\dfrac{3}{2}\log\left(\dfrac{\mathcal{M}}{12}+\dfrac{1}{12}\right)=-\dfrac{3}{2}\log\left(G E\right)+constant.
\end{equation}
which is independent of angular momentum.
In the following we consider the effect of correction term to the phase transition. So, by using the first law of thermodynamics the free energy becomes
\begin{equation}
F_{FSC}=\dfrac{\pi^2 T^2}{2G\Omega^2}\left(\bar{\sigma}+\dfrac{\alpha H}{\mu}+\dfrac{F}{m^2}+\dfrac{\Omega}{\mu}\right)-3T\log\left(\dfrac{10}{2\pi T}\right),
\end{equation}
by comparing the free energies, the conditions for phase transition change as follow
\begin{align}
&\mathcal{M}\left(1+\dfrac{\Omega}{\mu\left(\bar{\sigma}+\dfrac{\alpha H}{\mu}+\dfrac{F}{m^2}\right)}\right)+\dfrac{24 G T\log\left(\dfrac{10}{2\pi T}\right)}{\left(\bar{\sigma}+\dfrac{\alpha H}{\mu}+\dfrac{F}{m^2}\right)}<1,\;\;\;\;\textit{FSC is preferred}\nonumber\\
&\mathcal{M}\left(1+\dfrac{\Omega}{\mu\left(\bar{\sigma}+\dfrac{\alpha H}{\mu}+\dfrac{F}{m^2}\right)}\right)+\dfrac{24 G T\log\left(\dfrac{10}{2\pi T}\right)}{\left(\bar{\sigma}+\dfrac{\alpha H}{\mu}+\dfrac{F}{m^2}\right)}=1,\;\;\;\;\textit{FSC and HFS coexist}\nonumber\\
&\mathcal{M}\left(1+\dfrac{\Omega}{\mu\left(\bar{\sigma}+\dfrac{\alpha H}{\mu}+\dfrac{F}{m^2}\right)}\right)+\dfrac{24 G T\log\left(\dfrac{10}{2\pi T}\right)}{\left(\bar{\sigma}+\dfrac{\alpha H}{\mu}+\dfrac{F}{m^2}\right)}<1,\;\;\;\;\textit{HFS is preferred},
\end{align}
then the critical temperature becomes
\begin{equation}
T_{c}=\dfrac{\Omega}{2\pi}\dfrac{1}{\sqrt{1+\dfrac{\Omega}{\mu\left(\bar{\sigma}+\dfrac{\alpha H}{\mu}+\dfrac{F}{m^2}\right)}}}\sqrt{1+\dfrac{24GT\log\left(\dfrac{2\pi T}{10}\right)}{\left(\bar{\sigma}+\dfrac{\alpha H}{\mu}+\dfrac{F}{m^2}\right)}}=T_{c,0}\sqrt{1+\dfrac{24GT\log\left(\dfrac{2\pi T}{10}\right)}{\left(\bar{\sigma}+\dfrac{\alpha H}{\mu}+\dfrac{F}{m^2}\right)}},
\end{equation}
if we assume the second term to be small, then
\begin{equation}
T_{c} \sim T_{c,0}+T_{c,0}\dfrac{12GT\log\left(\dfrac{2\pi T}{10}\right)}{\left(\bar{\sigma}+\dfrac{\alpha H}{\mu}+\dfrac{F}{m^2}\right)}.
\end{equation}
So, the correction to entropy causes a positive shift to the critical temperature, i.e, the phase transition occurs in higher temperatures.

\section{Phase Transition in EGMG}\label{PTEGMG}
Exotic general massive gravity is the exotic version of NMG which is a parity-odd theory describing a propagating massive spin-2 fields in three dimensions. The Lagrangian of the theory in terms of auxiliary fields $f$ and $h$ is \cite{Ozkan:2018cxj}
\begin{align}
L_{EGMG}=&-\dfrac{1}{m^2}[f.R(\omega)+\dfrac{1}{6m^4}f.f\times f-\dfrac{1}{2 m^2}f.D(\omega)f+\dfrac{\nu}{2}f.e\times e-m^2h.T(\omega)+\nonumber\\
&\dfrac{(\nu-m^2)}{2}\left(\omega.d\omega+\dfrac{1}{3}\omega.\omega\times \omega\right)+\dfrac{\nu m^4}{3\mu}e.e\times e],\;\;\;\;\nu=\dfrac{1}{l^2}-\dfrac{m^4}{\mu^2}.
\end{align}
The field equation of the theory in the metric formalism is obtained as follows \cite{Ozkan:2018cxj}
\begin{equation}\label{eqq1}
R_{\mu \nu}-\dfrac{1}{2}g_{\mu \nu}R+\Lambda g_{\mu \nu}+\dfrac{1}{\mu}C_{\mu \nu}-\dfrac{1}{m^{2}}H_{\mu \nu}+\dfrac{1}{m^{4}}L_{\mu \nu}= 0,
\end{equation}
where
\begin{equation}\label{eqq2}
H_{\mu \nu}=\epsilon_{\mu}^{\alpha \beta}\nabla_{\alpha}C_{\nu \beta},\hspace{0.5cm}L_{\mu \nu}=\dfrac{1}{2}\epsilon_{\mu}^{\alpha \beta}\epsilon_{\nu}^{\gamma \sigma}C_{\alpha \gamma}C_{\beta \sigma}.
\end{equation}
$C_{\mu \nu}$ is the Cotton tensor which defined in (\ref{eq2}). $\mu$ and $m$ are mass parameters and $H_{\mu \nu}$ and $L_{\mu \nu}$ are traceless and symmetric tensors. This theory supplements the Einstein equations with a term that contains up to 3rd of the metric and is built with combinations and derivatives of the Cotton tensor \cite{Ozkan:2018cxj}, \cite{Afshar:2019npk}. The ADT charges \cite{Mann:2018vum}, the conserved charges by using of a variant of the ADT method \cite{Bergshoeff:2019rdb}, critical points \cite{Giribet:2019vbj} and the local causality of the theory studied in \cite{Kilicarslan:2019ply}.\\
The metric (\ref{cosmic1}) is a vacuum solution ($\Lambda_0=0$) to the field equation (\ref{eqq1}) and (\ref{eqq2}).
Here, we consider the phase transition between two spacetimes in the exotic general massive gravity.
The conserved charges of exotic general massive gravity based on a variant of the ADT method for Chern-Simons-like theories and for FSC metric obatin as follows \cite{Bergshoeff:2019rdb}
\begin{equation}
Q_{EGMG}(\xi)=-\dfrac{m^2}{\mu}Q_{GR}(\xi)+\left(1+\dfrac{m^2}{\mu^2}\right)Q_{EG}(\xi),
\end{equation}
where
\begin{equation}
Q_{EG}(\partial_t)=Q_{GR}(\partial_\phi)=-\dfrac{\mathcal{L}}{4G},\;\;\;\;Q_{EG}(\partial_\phi)=Q_{GR}(\partial_t)=\dfrac{\mathcal{M}}{8G}.
\end{equation}
The Bekenstein-Hawking entropy becomes \cite{Bergshoeff:2019rdb}
\begin{equation}\label{eqSS}
S=-\dfrac{m^2}{\mu}S_{GR}+\left(1+\dfrac{m^2}{\mu^2}\right)S_{EG}=\dfrac{\pi}{2G}\left[\dfrac{m^2}{\mu}\dfrac{\mathcal{L}}{\sqrt{\mathcal{M}}}+\left(1+\dfrac{m^2}{\mu^2}\right)\sqrt{\mathcal{M}}\right]=\dfrac{\pi^2 T}{G \Omega^2}\left[\dfrac{m^2}{\mu}+\left(1+\dfrac{m^2}{\mu^2}\right)\Omega\right].
\end{equation}
So, the free energy by integrating the first law of thermodynamics (\ref{eqflaw}) is obtained as follows
\begin{equation}\label{eqfree1}
F_{FSC}=-\dfrac{\mathcal{M}}{8G}\left[\dfrac{m^2}{\mu}+\left(1+\dfrac{m^2}{\mu^2}\right)\Omega\right]=-\dfrac{\pi^2 T^2}{2G \Omega^2}\left[\dfrac{m^2}{\mu}+\left(1+\dfrac{m^2}{\mu^2}\right)\Omega\right],
\end{equation}
and the free energy of HFS ($\mathcal{M}=1, \Omega=0$) is
\begin{equation}\label{eqfree2}
F_{HFS}=-\dfrac{1}{8G}\dfrac{m^2}{\mu}.
\end{equation}
By comparing the free energies (\ref{eqfree1}) and (\ref{eqfree2}) we come to the following conditions for the phase transition
\begin{align}
&\mathcal{M}\left[1+\dfrac{\Omega}{\mu}\left(1+\dfrac{\mu^2}{m^2}\right)\right]>1,\;\;\;\;\textit{FSC is preferred}\nonumber\\
&\mathcal{M}\left[1+\dfrac{\Omega}{\mu}\left(1+\dfrac{\mu^2}{m^2}\right)\right]=1\;\;\;\;\textit{FSC and HFS coexist}\nonumber\\
&\mathcal{M}\left[1+\dfrac{\Omega}{\mu}\left(1+\dfrac{\mu^2}{m^2}\right)\right]<1\;\;\;\;\textit{HFS is preferred}.\nonumber\\
\end{align}
If $\mu>0$, $\Omega>0$, and $ \Omega/\mu\left(1+\mu^2/m^2\right)<1 $, then the FSC is the stabel spacetime. Also, in the case $\mu>0$ and $\Omega<0$, there is a condition for stability of FSC.

The critical temprature is
\begin{equation}
T_{c}=\dfrac{\Omega}{2\pi}\dfrac{1}{\sqrt{1+\dfrac{\Omega}{\mu}\left(1+\dfrac{\mu^2}{m^2}\right)}}.
\end{equation}
Similar to GMMG, at sufficiently high temperatures, HFS melts into FSC. Conversely,
at sufficiently high angular velocity, HFS is more stable and FSC tunnels to HFS.\\
Further more, by using (\ref{eqc21}) and (\ref{eqSS}) one can find that the FSC spacetime in EGMG is also localy stable, i.e, $C=S>0$.
\subsection{Logarithmic Correction to Entropy}
 The AdS$_{3}$ central charges of the asymptotic conformal symmetry algebra according to the computation of \cite{Bergshoeff:2019rdb} gives
\begin{equation}\label{centlag}
c_{\pm}=\dfrac{3l}{2G}\left[-\dfrac{m^2 l}{\mu}\pm\left(1+\dfrac{m^2}{\mu^2}-\dfrac{1}{l^2 m^2}\right)\right],
\end{equation}
if taking the limit $\mu\rightarrow \infty$ and then taking the limit $m\rightarrow \infty$, one obtains $c_{\pm}=3l/2G$, which agrees with the central charges of the Exotic Gravity \cite{Bergshoeff:2019rdb}.
The BMS$_3$ central charges correspond to the asymptotic flat spacetimes are as follows
\begin{equation}\label{centseq}
c_{M}=\lim_{l\rightarrow \infty}\dfrac{1}{{l^2}}(c_{+}+c_{-})=-\dfrac{3m^2}{\mu G},\;\;\;\;c_{L}=\lim_{l\rightarrow \infty}\dfrac{1}{{l}} (c_{+}-c_{-})=\dfrac{3}{G}\left[1+\dfrac{m^2}{\mu^2}\right].
\end{equation}

The conformal weights using the BMS$_{3}$ central charges (\ref{centseq}) are
\begin{align}\label{eqcww}
&h_{M}=Q_{EGMG}(\partial_{t})+\dfrac{c_{M}}{24}=-\dfrac{m^2\mathcal{M}}{8\mu G}-\left(1+\dfrac{m^2}{\mu^2}\right)\dfrac{\mathcal{L}}{4G}-\dfrac{m^2}{8\mu G},\nonumber\\
&h_{L}=Q_{EGMG}(\partial_{\phi})+\dfrac{c_{L}}{24}=\dfrac{m^2}{\mu}\dfrac{\mathcal{L}}{4G}+\left(1+\dfrac{m^2}{\mu^2}\right)\dfrac{\mathcal{M}}{8G}+\dfrac{1}{8G}\left[1+\dfrac{m^2}{\mu^2}\right].
\end{align}
Finally, by using equations (\ref{eqents}), (\ref{eqcww}) and large $\mathcal{M}$, $\mathcal{L}$, one can obtain the entropy as follows
\begin{equation}\label{eqsen}
S^{(0)}=\dfrac{\pi}{2G}\sqrt{\mathcal{M}}\left[1+\dfrac{m^2}{\mu^2}-\dfrac{m^2}{\mu \Omega}\right],
\end{equation}
as can be seen, the above result for $S^{(0)}$ is the same as Bekenstein-Hawking entropy (\ref{eqSS}). The correction to entropy is
\begin{equation}
S^{(1)}=-\dfrac{3}{2}\log\left(\dfrac{m^2 \mu \mathcal{M}+2\mathcal{L}(\mu^2+m^2)}{12\mu m^2 }\right)=-\dfrac{3}{2}\log\left(G Q_{EGMG}(\partial_{t})\right)+constant
\end{equation}
which is independent of parameters of theory and angular momentum.
In the following we consider the effect of correction term to the phase transition. So, by using the first law of thermodynamics the free energy becomes
\begin{equation}
F_{FSC}=-\dfrac{\pi^2 T^2}{2G \Omega^2}\left[\dfrac{m^2}{\mu}+\left(1+\dfrac{m^2}{\mu^2}\right)\Omega\right]-3T\log\left(\dfrac{10m\sqrt{\mu}}{2\pi T}\right),
\end{equation}
by comparing the free energies, the conditions for phase transition change as follow
\begin{align}
&\mathcal{M}\left[1+\dfrac{\Omega}{\mu}\left(1+\dfrac{\mu^2}{m^2}\right)\right]+\dfrac{24 G\mu T}{m^2}\log\left(\dfrac{10m\sqrt{\mu}}{2\pi T}\right)>1,\;\;\;\;\textit{FSC is preferred}\nonumber\\
&\mathcal{M}\left[1+\dfrac{\Omega}{\mu}\left(1+\dfrac{\mu^2}{m^2}\right)\right]+\dfrac{24 G\mu T}{m^2}\log\left(\dfrac{10m\sqrt{\mu}}{2\pi T}\right)=1,\;\;\;\;\textit{FSC and HFS coexist}\nonumber\\
&\mathcal{M}\left[1+\dfrac{\Omega}{\mu}\left(1+\dfrac{\mu^2}{m^2}\right)\right]+\dfrac{24 G\mu T}{m^2}\log\left(\dfrac{10m\sqrt{\mu}}{2\pi T}\right)<1,\;\;\;\;\textit{HFS is preferred}
\end{align}
then the critical temperature becomes
\begin{equation}
T_{c}=\dfrac{\Omega}{2\pi}\dfrac{1}{\sqrt{1+\dfrac{\Omega}{\mu}\left(1+\dfrac{\mu^2}{m^2}\right)}}\sqrt{1+\dfrac{24 G\mu T}{m^2}\log\left(\dfrac{2\pi T}{10m\sqrt{\mu}}\right)}=T_{c,0}\sqrt{1+\dfrac{24 G\mu T}{m^2}\log\left(\dfrac{2\pi T}{10m\sqrt{\mu}}\right)}
\end{equation}
if assume the second term to be small, then
\begin{equation}
T_{c} \sim T_{c,0}+\dfrac{12 G\mu T_{c,0} T}{m^2}\log\left(\dfrac{2\pi T}{10m\sqrt{\mu}}\right).
\end{equation}
So, the correction to entropy causes a positive shift to the critical temperature, i.e, the phase transition occurs in higher temperatures.

\section{Conclusion}\label{conclusion}
In the present work, we studied the phase transition between hot flat spacetime and flat cosmological spacetime, in the framework of GMMG and EGMG models. To study the phase transition, we obtained the thermodynamics quantities, and free energy by using the first law of thermodynamics. By comparing the free energies of FSC and HFS, we determined the thermodynamic stability of spacetime for both theories. We conclude that at high temperatures and low rotations the phase transition occurs from HFS to FSC and preferred spacetime is FSC. Considering the correction term for entropy, we see that the phase transition occurs at a higher temperature. Then, by obtaining the BMS$_{3}$ central charges and using Cardy formula we checked the consistency with the thermodynamic entropy.  Regarding EGMG, we have considered two types of conserved charges.  First, by using of the conserved charges which have been obtained according to a type of the ADT method to Chern-Simons-like theories of gravity, and the other one, using the conserved charges which obtained according to the metric formulation of ADT approach. In both types the phase transition occurs but in different conditions and temperatures. As can be seen from the appendix \ref{appa}, the conserved charges obtained by using of the field equations of EGMG (similary for GMMG) for the metric (\ref{cosmic1}) are the same as those one in TMG \cite{Bagchi:2013lma}. This means that the higher derivative terms in the field equations of EGMG and GMMG do not contributions to the conserved charges of the metric.

\appendix

\section{Phase Transition in EGMG Using Second Kinds of Conserved Charges}\label{appa}
Here, we want to investigate the phase transition using the conserved charges of \cite{Mann:2018vum} which has been obtained by using of linearzed field equations (\ref{eqq1}) around a maximally symmetric background and using ADT method \cite{ADTs}. The ADT method gives a conserved charge for each Killing vector of the background, expressed in terms of an integral over the perturbed field equations near spatial infinity \cite{ADTs}.
So, the conserved quantities for solution (\ref{cosmic1}) with background metric (\ref{Bmetric}) become \cite{Mann:2018vum}
\begin{equation}\label{eqE1}
E=\dfrac{\mathcal{M}}{8G},
\end{equation}
\begin{equation}\label{eqJ1}
 J=\dfrac{\mathcal{L}}{4G}-\dfrac{\mathcal{M}}{8\mu G},
\end{equation}
\begin{equation}\label{eqent2}
S=\dfrac{\pi}{2G}\left(\dfrac{\mathcal{L}}{\sqrt{\mathcal{M}}}-\dfrac{\sqrt{\mathcal{M}}}{\mu}\right).
\end{equation}
By looking at field equations (\ref{eqq1}) can be seen, even though the tensors $G_{\mu \nu}$ and $C_{\mu \nu}$ vanish, we saw that they do contribute non-trivially to the charges. Also, the parity odd term $H$ and parity even $L$ do not contribute to the charges.\\
By using of equation (\ref{eqflaw}) for the first law of thermodynamics we obtain free energy for FSC as follows
\begin{equation}
F_{FSC}=-\dfrac{\mathcal{M}}{8G}\left(1+\dfrac{\Omega}{\mu}\right),
\end{equation}
while for flat space we have
\begin{equation}
F_{HFS}=-\dfrac{1}{8G}.
\end{equation}
Comparing the free energies and assuming $\mu>0$, give us such conditions for phase transition
\begin{align}
&\mathcal{M}\left(1+\dfrac{\Omega}{\mu}\right)>1,\;\;\;\;\;\textit{FSC is preferred}\nonumber\\
&\mathcal{M}\left(1+\dfrac{\Omega}{\mu}\right)=1,\;\;\;\;\;\textit{FSC and HFS coexist},\nonumber\\
&\mathcal{M}\left(1+\dfrac{\Omega}{\mu}\right)<1,\;\;\;\;\;\textit{HFS is preferred}
\end{align}
The phase transition occurs at the critical temperature
\begin{equation}
T_{c}=\dfrac{\Omega}{2\pi}\dfrac{1}{\sqrt{1+\dfrac{\Omega}{\mu}}},
\end{equation}
at sufficiently high temperatures, HFS tunnels into FSC and localy is stable.

\subsection{Logarithmic Correction to Entropy}

Here, we consider the AdS$_3$ central charges of EGMG according to \cite{Giribet:2019vbj} and \cite{Giribet:2019wme}. So, we have
\begin{equation}\label{eqEGMGcen}
c_{\pm}=\dfrac{3l}{2G}\left(1-\dfrac{1}{m^2 l^2}\mp\dfrac{1}{\mu l}\right).
\end{equation}
It should be note these central charges are different from (\ref{centlag}) \cite{Bergshoeff:2019rdb}.
By considering the above result for central charges of EGMG and takes the limit $\mu\rightarrow \infty$, $m\rightarrow\infty$ of that one obtains $c_{\pm}=3l/2G$ which is the Brown-Henneaux central charges of Einstein gravity \cite{Brown:1986nw}.
By using (\ref{eqcentral}) and (\ref{eqEGMGcen}) one can obtain BMS$_3$ central charges as follows
\begin{equation}\label{eqcenbms}
c_{L}=-\dfrac{3}{\mu G},\;\;\;\;\; c_{M}=\dfrac{3}{G}.
\end{equation}
 According to equations (\ref{eqE1}) and (\ref{eqJ1}) the conformal weights become
\begin{align}
&h_{M}=\lim_{l\rightarrow \infty}\dfrac{h+\bar{h}}{l}=\dfrac{1}{8G}\left(\mathcal{M}+1\right),\nonumber\\
&h_{L}=\lim_{l\rightarrow \infty}(h-\bar{h})=\dfrac{1}{4G}\left[\mathcal{L}-\dfrac{\mathcal{M}}{2\mu}-\dfrac{1}{2\mu}\right],
\end{align}
by using equation (\ref{eqents}) one can obtain the entropy and its correction as follows \cite{Carlip:2000nv}
\begin{equation}\label{eqentr}
S=\dfrac{\pi}{2\mu G}\dfrac{1}{\sqrt{\mathcal{M}+1}}\left[-\mathcal{M}-1+\mu \mathcal{L}\right]-\dfrac{3}{2}\log\left(\dfrac{\mathcal{M}+1}{12}\right).
\end{equation}
In the case $\mathcal{M},\mathcal{L}\gg 1$, the zero term of equations (\ref{eqentr}) and (\ref{eqent2}) are the same, and the correction term is the same as Einstien gravity.\\
In the following we consider the effect of correction term to the phase transition. So, by using the first law of thermodynamics the free energy becomes
\begin{equation}
F_{FSC}=-\dfrac{\pi^2 T^2}{2G\Omega^2}\left(1+\dfrac{\Omega}{\mu}\right)-3T\log\left(\dfrac{10}{2\pi T}\right),
\end{equation}
by comparing the free energies, the conditions for phase transition change as follow
\begin{align}
&\mathcal{M}\left(1+\dfrac{\Omega}{\mu}\right)+24 G T\log\left(\dfrac{10}{2\pi T}\right)>1,\;\;\;\;\textit{FSC is preferred}\nonumber\\
&\mathcal{M}\left(1+\dfrac{\Omega}{\mu}\right)+24 G T\log\left(\dfrac{10}{2\pi T}\right)=1,\;\;\;\;\textit{FSC and HFS coexist}\nonumber\\
&\mathcal{M}\left(1+\dfrac{\Omega}{\mu}\right)+24 G T\log\left(\dfrac{10}{2\pi T}\right)<1,\;\;\;\;\textit{HFS is preferred}
\end{align}
then the critical temperature becomes
\begin{equation}
T_{c}=\dfrac{\Omega}{2\pi}\dfrac{1}{\sqrt{1+\dfrac{\Omega}{\mu}}}\sqrt{1+24GT\log\left(\dfrac{2\pi T}{10}\right)}=T_{c,0}\sqrt{1+24GT\log\left(\dfrac{2\pi T}{10}\right)}
\end{equation}
if assume the second term to be small, then
\begin{equation}
T_{c} \sim T_{c,0}+12GT_{c,0}T\log\left(\dfrac{2\pi T}{10}\right).
\end{equation}
So, the correction to entropy causes a positive shift to the critical temperature, i.e, the phase transition occurs in higher temperatures.
 
\end{document}